\documentclass[epj,twocolumn]{webofc}
\usepackage[varg]{txfonts}   
\woctitle{Transversity 2014. Fourth International Workshop on Transverse Polarisation Phenomena in Hard Processes. 9-13 June, 2014 - Chia, Cagliari, Italy}

\usepackage{pdfsync}
\usepackage{bm}
\usepackage{slashed}
\usepackage{graphicx}
\usepackage{subfigure}
\usepackage[svgnames]{xcolor}
\usepackage{placeins}
\usepackage{xspace}

\def\LEPT{\texttt{LEPTO}\xspace}
\def\MLEPT{\texttt{\lowercase{m}LEPTO}\xspace}

\def\vfR{$\varphi_R$\xspace}
\def\fR{\varphi_R}
\def\fS{\varphi_S}
\def\vfT{$\varphi_T$\xspace}
\def\fT{\varphi_T}
\def\kT{\vect{k}_T}
\def\kt{k_T}

\def\Pp#1{\vect{P}_{#1\perp}}
\def\PT#1{\vect{P}_{#1T}}
\newcommand{\vT}{\vect{P}_T}
\newcommand{\vect}[1]{\boldsymbol{#1}}
\newcommand{\bfk}{\vect{k}}
\newcommand{\bfq}{\vect{q}}

\newcommand{\bfS}{\vect{S}_{_T}}

\newcommand{\al}[1]{\begin{align} #1 \end{align}}
\newcommand{\non}{\nonumber}
\newcommand{\vf}{\varphi}

\newcommand{\Ge}{\mathrm{GeV}}
\newcommand{\Gs}{\mathrm{GeV}^2}
\newcommand{\ImS}{0.8\columnwidth}
\newcommand{\Eq}[1]{Eq.~(\ref{#1})}

\begin{document}

\title{Sivers Effect in Dihadron Electroproduction}

\author{Aram~Kotzinian\inst{1,2}\fnsep\thanks{Speaker} \and
        Hrayr~H.~Matevosyan\inst{3} \and
        Anthony~W.~Thomas\inst{3}}

\institute{Yerevan Physics Institute, 2 Alikhanyan Brothers St., 375036 Yerevan, Armenia
\and INFN, Sezione di Torino, 10125 Torino, Italy
\and ARC Centre of Excellence for Particle Physics at the Tera-scale,
\\CSSM, School of Chemistry and Physics, The University of Adelaide, Adelaide SA 5005, Australia
\\ http://www.physics.adelaide.edu.au/cssm }


\abstract{%
The Sivers effect in polarized SIDIS can be measured in two hadron production as sine modulations  involving the azimuthal angles \vfT~and \vfR~of both the total and the relative transverse momenta of the hadron pair, complementary to the conventional single hadron studies. In this talk we briefly present the results obtained in our recent work~\cite{Kotzinian:2014lsa} and~\cite{Kotzinian:2014gza}. We also present the leading order parton model expression for the two hadron SIDIS cross section for different choices of the relative transverse momentum that dismiss the seeming contradiction of our results with  previous work. Finally, we show the numerical predictions for the corresponding single spin asymmetries in the kinematics of COMPASS experiment obtained using the modified version of the \LEPT Monte Carlo event generator that includes the Sivers effect.
}
\vspace{-1.5cm}
\maketitle
\vspace{-0.5cm}
\section{Introduction}
\label{SEC_INTRO}

One of the most interesting phenomena related to internal 3D structure of polarized nucleons is the Sivers effect~\cite{Sivers:1989cc}. The corresponding SSA in one hadron SIDIS have been measured in HERMES~\cite{Airapetian:2009ae}, COMPASS~\cite{Adolph:2012sp}, and JLab~\cite{Qian:2011py} experiments and allow to extract the Sivers PDF, e.g.~\cite{Anselmino:2005nn}.

Recently, the two hadron production in polarized SIDIS (2h SIDIS) and in the jets from $e^+e^-$ annihilation attracted a lot of interest both in experimental~\cite{Airapetian:2008sk, Vossen:2011fk, Adolph:2014fjw} and theoretical~\cite{Artru:1995zu, Bianconi:1999cd, Bacchetta:2003vn} studies. The main accent has been on studies of transversity PDFs and so called interference fragmentation functions.

Here we present the phenomenology of Sivers effect in 2h SIDIS. We also discuss the different choices for the relative transverse momentum of the hadrons pair and their implications on the observable Sivers asymmetry.

\vspace{-0.5cm}
\section{The Sivers effect in 2h SIDIS}
\label{SEC_SIV_2H}
%
\begin{figure}[tbh]
\begin{center}
\includegraphics[width=0.9\columnwidth]{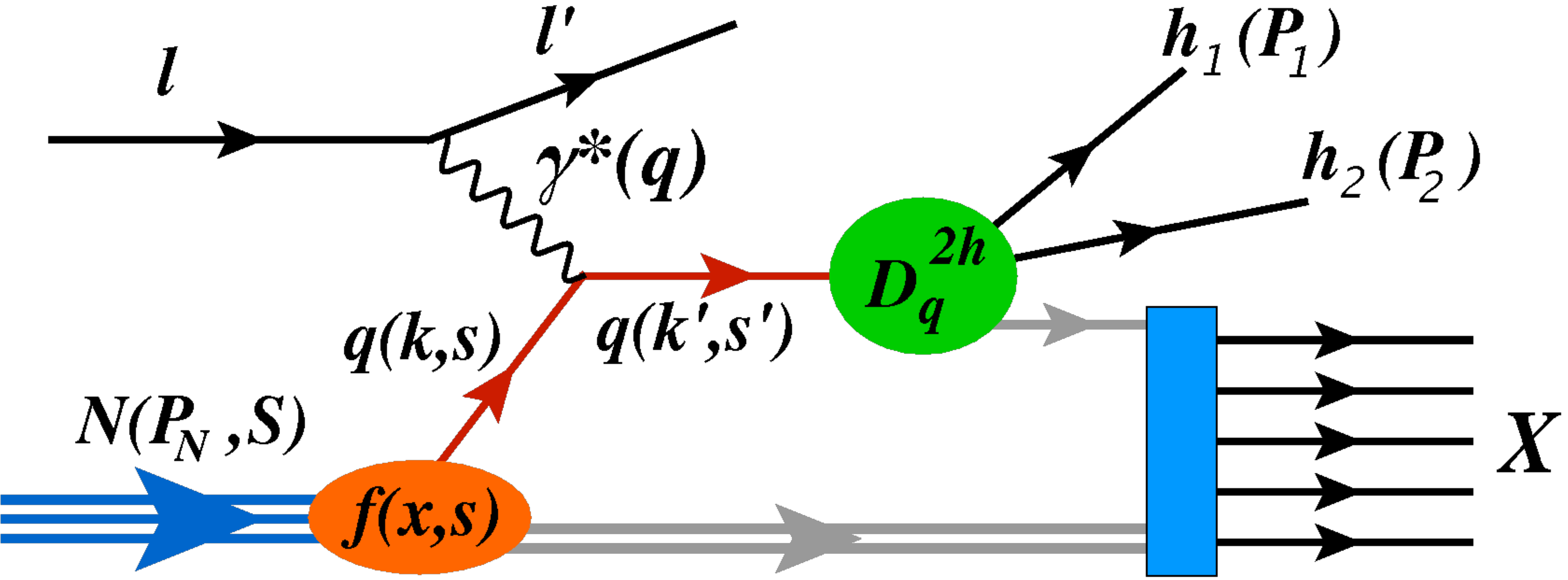}
\caption{Two hadron production in the current fragmentation region of SIDIS.}
\label{FIG_DIHADRON_CFR}
\end{center}
\end{figure}
%
\subsection{Cross section expression}
\label{SUB_SEC_XSEC_H1_H2}

\begin{figure}[h]
\begin{center}
\includegraphics[width=0.9\columnwidth]{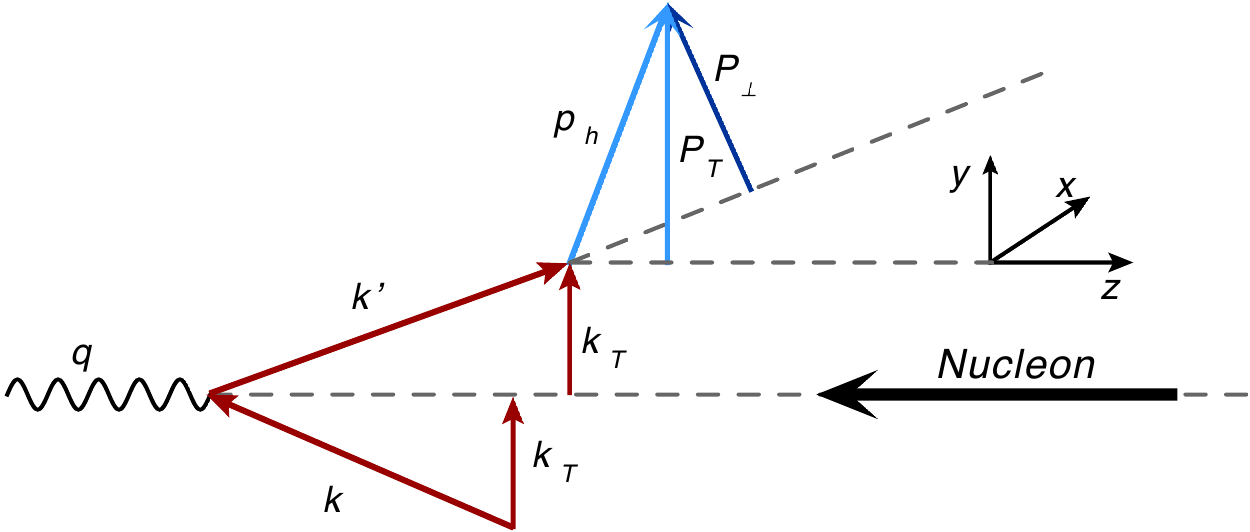}
\caption{$\gamma^*-N$ center of mass frame.}
\label{FIG_GAMMA-N_FRAME}
\end{center}
\end{figure}

 The process of two hadron electro-production in the current fragmentation region of SIDIS is schematically depicted in Fig.~\ref{FIG_DIHADRON_CFR}. In this work we adopt the $\gamma^*-N$ center of mass frame, where the $z$ axis is along the direction of the virtual photon momentum $\bfq$. The $x$-$z$ plane is the lepton scattering plane. In this frame, we will define the transverse components of the momenta with respect to the $z$ axis with subscript $_T$ and the transverse momenta with respect to the fragmenting quark's direction with subscript $_\perp$, as demonstrated in Fig.~\ref{FIG_GAMMA-N_FRAME}. We use that standard SIDIS kinematical variables as described in Ref.~\cite{Anselmino:2005nn}.

 It is easy to see that in this system $\vect{k}'_T=\vect{k}_T$ and (similar to single hadron SIDIS~\cite{Anselmino:2005nn}) in the DIS limit for small transverse momenta of the produced hadrons
\al
{
\label{EQ_PT}
\Pp{1} \approx \PT{1} - z_1 \kT,\;\;
\Pp{2} \approx \PT{2} - z_2 \kT.
}

We assume that the cross section of the process of interest factorizes into convolutions of transverse polarization dependent TMD PDFs, $f_\uparrow^q(x, \vect{k}_T)$, unpolarized dihadron fragmentation functions (DiFFs), $D_q^{h_1h_2}(z_1, z_2, \vect{P}_{1\perp}, \vect{P}_{2\perp})$ and a hard lepton-quark scattering cross section.
The polarized PDF  $f_\uparrow^q$ is composed of contributions from the unpolarized $f_1^q(x,k_T)$ and the Sivers term describing the correlation between the active quark's transverse momentum, $\vect{k}_T$ and the transverse polarization of the nucleon, $\mathbf{S}_T$,
\al{
f_\uparrow^q(x,\vect{k}_T)=f_1^q(x,k_T)+\frac{[\bfS \times \bfk_T]_3}{M}f_{1T}^{\perp q}(x,k_T),
}
where $f_{1T}^{\perp q}(x,k_T)$ denotes the Sivers PDF,  $M$ is the nucleon mass and the subscript $3$ denotes the $z$ component of the vector. Then, using \Eq{EQ_PT}, the 2h SIDIS cross section can be separated into the usual unpolarized part, $\sigma_U$, and a spin-dependent part induced by the Sivers effect, $\sigma_S$:
\al{
\non
& \frac{d\sigma^{h_1h_2}}{dx\, d{Q^2}\, d{\fS}\, dz_1\, dz_2\, d^2\PT{1}\, d^2\PT{2}} = C(x,Q^2) \left({\sigma_U} + {\sigma_{S}}\right),
\\ \non
& \sigma_U = \sum_q e_q^2 \int d^2 \kT d^2 \vect{P}_{1\perp} d^2 \vect{P}_{2\perp}\delta^2\left(\PT{1} - \Pp{1} - z_1 \kT\right)
\\ \non
& \times \delta^2\left(\PT{2} - \Pp{2} - z_2 \kT\right)f_1^q\left(x,k\right)\
D_q^{h_1h_2}\left(z_1,z_2,\vect{P}_{1\perp},\vect{P}_{2\perp}\right),
\\ \non
&\sigma_{S} = \sum_q e_q^2 \int d^2 \kT d^2 \vect{P}_{1\perp} d^2 \vect{P}_{2\perp}
\delta^2\left(\PT{1} - \Pp{1} - z_1 \kT\right)
\\ \non
& \times \delta^2\left(\PT{2} - \Pp{2} - z_2 \kT\right)\frac{[\bfS \times \bfk_T]_3}{M}
f_{1T}^{ \perp q}\left(x,k\right)\
\\
\label{EQ_2H_SIG_SIV}
& D_q^{h_1h_2}\left(z_1,z_2,\vect{P}_{1\perp},\vect{P}_{2\perp}\right),
}
where $C(x,Q^2) =  \frac{\alpha^2}{Q^4}(1+(1-y)^2)$ and $\alpha$ is the fine-structure coupling constant.

  It is easy to see using rotational and parity invariance, that the most general dependence of $\sigma_S$ on the azimuthal angles $\vf_{1,2}$ (of the transverse momenta $\vect{P}_{1,2T}$) and~$\fS$ is given by two "Sivers-like" terms
\al{
\label{EQ_2H_SIV_GEN}
\sigma_{S}
= S_T\left(\sigma_1\frac{P_{1T}}{M}  \sin(\vf_1-\fS) \right.&
\\ \non
+
&\left. \sigma_2\frac{P_{2T}}{M} \sin(\vf_2-\fS)\right),
}
where $\sigma_{S}, \sigma_1$ and $\sigma_2$ depend on $x,Q^2,z_1,z_2,P_{1T},P_{2T}$ and $\vect{P}_{1T}\cdot\vect{P}_{2T}$ (or $\cos(\vf_1-\vf_2)$).

The explicit expressions for structure functions $\sigma_U$, $\sigma_1$ and $\sigma_2$ are presented in Ref.~\cite{Kotzinian:2014gza} using Gaussian parameterizations for PDFs, similar to Ref.~\cite{Anselmino:2005nn}, and the following form for DiFFs:
\al{
\label{EQ_UNPOL_DIFF}
&D_{1q}^{h_1h_2}\left(z_1,z_2,P_{1\perp},P_{2\perp},\vect{P}_{1\perp}\cdot\vect{P}_{2\perp}\right)=
\\&\non
D_{1q}^{h_1h_2}(z_1,z_2)\frac{1}{\pi^2 \nu_1^2 \nu_2^2} \, e^{-P_{1\perp}^2/\nu_1^2-P_{2\perp}^2/\nu_2^2}
\left(1+c\Pp{1} \cdot \Pp{2}\right),
}
where the term ${c \Pp{1} \cdot \Pp{2}}$ takes into account the experimentally established transverse momentum correlations  (see, e.g. Ref.~\cite{Arneodo:1986yc}, where $a^{-2}$ coefficient is used instead of $c$) in unpolarized 2h SIDIS.

\subsection{Choice of the relative transverse momentum}

 It is often convenient to use linear combinations of $\PT{1}$ and $\PT{2}$ as independent transverse momentum variables. One vector is chosen as the transverse component of the pair's total momentum,
\al{
&
 \vect{P}=\vect{P}_1+\vect{P}_2,
\\
&
\vT=\vect{P}_{1T}+\vect{P}_{2T}.
}

 For the second vector, the following choices have been considered
 \begin{enumerate}
  \item In Refs.~\cite{Vossen:2011fk,Kotzinian:2014lsa,Kotzinian:2014gza} -- the transverse component of the relative momentum
\al{
&
\vect{R}=\frac{1}{2}\left(\vect{P}_{1}-\vect{P}_{2}\right),
\\
&
\vect{R}_T=\frac{1}{2}\left(\vect{P}_{1T}-\vect{P}_{2T}\right).
}
  \item In Refs.~\cite{Artru:1995zu,Adolph:2014fjw} -- the weighted difference
\al{
\vect{R}_{T,A}=\xi_2 \vect{P}_{1T}-\xi_1 \vect{P}_{2T},
}
where
\al{
\xi_i=z_i/z,\,\; \;
 z=z_1+z_2.
}
  \item In Ref.~\cite{Vossen:2011fk,Bianconi:1999cd,Airapetian:2008sk} -- the transverse to $\gamma^*$-direction component of the transverse to pair total momentum, $\vect{P}$, component of $\vect{R}$:
\al{
  \vect{R}_{T,P}=\vect{R}_{P}-\left(\vect{R}_{P}\cdot\vect{\hat q}\right)\vect{\hat q},
}
where
\al{
\vect{R}_{P}=\vect{R}-\left(\vect{R}\cdot\vect{\hat P}\right)\vect{\hat P},
}
and $\vect{\hat v}$ denotes the unit vector in the direction of vector $\vect{v}$.
\end{enumerate}

The explicit calculations in the leading order approximation where we neglect mass terms and for small transverse momenta compared to large $Q$
\al
{
\vect{P}_i^2 = E_i^2 -m_i^2 \approx z_{i}^2 \vect{k}'^2,
}
then
\al
{
\vect{R}_P \approx \vect{R} -\frac{z_{1}^2 - z_{2}^2}{(z_{1}+z_{2})^2} \frac{\vect{P}}{2}
= \xi_2 \vect{P}_1- \xi_1 \vect{P}_2,
}
and
\al{
\label{EQ_R_TAP}
\vect{R}_{T,P}\approx\vect{R}_{T,A}.
 }

It is useful to define the total and relative transverse momenta also in terms of transverse momenta acquired in fragmentation:
\al{
\non
&\vect{P}_{\perp} \doteq \vect{P}_{1,\perp} + \vect{P}_{2,\perp}, \quad
\vect{R}_{\perp} \doteq \frac{1}{2}\left(\vect{P}_{1,\perp} - \vect{P}_{2,\perp}\right),
\\
&\vect{R}_{\perp,A} \doteq \xi_2\vect{P}_{1,\perp} - \xi_1\vect{P}_{2,\perp}.
\label{EQ_R_P_frag}
 }
 The remarkable feature of the $\vect{R}_{T,A}$ is that at the leading order approximations it is independent on the quark intrinsic transverse momentum, $\kT$, due to relations Eq.~(\ref{EQ_PT}):
\al{
\label{EQ_R_TA}
\vect{R}_{T,A}=\xi_2\left(\Pp{1}+z_1\kT\right)-\xi_1\left(\Pp{2}+z_2\kT\right)=\vect{R}_{\perp,A},
}
whereas
\al{
\label{EQ_R_T_P_T}
\vect{P}_{T}=\vect{P}_{\perp}+z\kT, \quad
\vect{R}_{T}=\vect{R}_{\perp}+\frac{1}{2}\left(z_1-z_2\right)\kT
}
depend on $\kT$.

\subsection{Cross section in terms of relative and total hadron momenta}

Let us first rewrite Eq.~(\ref{EQ_2H_SIG_SIV}) in terms of $\vect{P}_T$ and $\vect{R}_{T,A}$\footnote{For brevity,  in the following we only consider the variables related to transverse components of vectors and omit in r.h.s. $C(x,Q^2)\sum_q e_q^2$.}
\al{
\non
& \frac{d\sigma^{h_1h_2}}{d{\fS}\, d^2\vect{P}_T\, d^2\vect{R}_{T,A}} = \int d^2 \kT d^2 \vect{P}_{\perp} d^2 \vect{R}_{\perp,A}
\\ \non
&\times\delta^2\left(\xi_1\left(\vect{P}_T - \vect{P}_{\perp}\right)+\vect{R}_{T,A}-\vect{R}_{\perp,A}-z_1 \kT\right)
\\ \non
&\times\delta^2\left(\xi_2\left(\vect{P}_T - \vect{P}_{\perp}\right)-\vect{R}_{T,A}+\vect{R}_{\perp,A}-z_2 \kT\right)
\\
&\times\left(f_1^q\left(k\right)+\frac{[\bfS \times \bfk_T]_3}{M}f_{1T}^{ \perp q}\left(k\right)\right)
D_{q,A}^{h_1h_2}\left(\vect{P}_{\perp},\vect{R}_{\perp,A}\right),
\label{EQ_2H_SIG_SIV_RA}
}
where we defined
\al{
\non
D_{q,A}^{h_1h_2}\left(\vect{P}_{\perp},\vect{R}_{\perp,A}\right)  &
\doteq \\
&
D_{q}^{h_1h_2}\left(\xi_1\vect{P}_{\perp}+\vect{R}_{\perp,A},\xi_2\vect{P}_{\perp}-\vect{R}_{\perp,A}\right).
\label{EQ_DA}
}
Note that, the product of $\delta$-functions in Eq.(\ref{EQ_2H_SIG_SIV_RA}) can be written as
$\delta^2\left(\vect{P}_T - \vect{P}_{\perp}-z \kT\right)\delta^2\left(\vect{R}_{T,A} - \vect{R}_{\perp,A}\right)$.
Then, after integrating over $\vect{R}_{\perp,A}$, we can write Eq.(\ref{EQ_2H_SIG_SIV_RA}) as
\al{
\non
& \frac{d\sigma^{h_1h_2}}{d{\fS}\, d^2\vect{P}_T\, d^2\vect{R}_{T,A}} = \int d^2 \kT d^2 \vect{P}_{\perp}
\delta^2\left(\vect{P}_T - \vect{P}_{\perp}-z \kT\right)
\\
& \times \left(f_1^q\left(k\right)+\frac{[\bfS \times \bfk_T]_3}{M}f_{1T}^{ \perp q}\left(k\right)\right)
D_{q,A}^{h_1h_2}\left(\vect{P}_{\perp},\vect{R}_{T,A}\right).
\label{EQ_2H_SIG_SIV_RA_intRperp}
}
The $\vect{P}_T$-integrated cross section becomes
\al{
\non
& \frac{d\sigma^{h_1h_2}}{d{\fS}\, d^2\vect{R}_{T,A}} = \int d^2 \kT d^2 \vect{P}_{\perp}
\\
& \times \left(f_1^q\left(k\right)+\frac{[\bfS \times \bfk_T]_3}{M}f_{1T}^{ \perp q}\left(k\right)\right)
D_{q,A}^{h_1h_2}\left(\vect{P}_{\perp},\vect{R}_{T,A}\right)
\non\\
& = \int d^2 \kT f_1^q\left(k\right)
\int d^2 \vect{P}_{\perp}D_{q,A}^{h_1h_2}\left(\vect{P}_{\perp},\vect{R}_{T,A}\right)
\label{EQ_2H_SIG_SIV_RA_intPT},
}
and does not depend on target polarization. This is what we have expected for the second choice of relative transverse momentum, since for SIDIS it is note sensitive to quark intrinsic transverse momentum. This feature of Sivers effect in terms of the third choice was well known~\cite{Bianconi:1999cd} and can be easily understood by considering Eq.~(\ref{EQ_R_TAP}).

In our studies~\cite{Kotzinian:2014lsa,Kotzinian:2014gza} of the Sivers effect in 2h SIDIS we have chosen the first definition of the relative transverse momentum. As for the previous choice, we can rewrite the Eq.~(\ref{EQ_2H_SIG_SIV}) in terms of $\vect{P}_T$ and $\vect{R}_T$:
\al{
\non
& \frac{d\sigma^{h_1h_2}}{d{\fS}\, d^2\vect{P}_Td^2\vect{R}_T} =
\int d^2 \kT d^2 \vect{P}_{\perp} d^2 \vect{R}_{\perp}
\delta^2\left(\vect{P}_T - \vect{P}_{\perp}-z \kT\right)
 \non \\
& \times \delta^2\left(\vect{R}_T - \vect{R}_{\perp}-\frac{1}{2}(z_1-z_2) \kT\right)
 \non \\
& \times \left(f_1^q\left(k\right)+\frac{[\bfS \times \bfk_T]_3}{M}f_{1T}^{ \perp q}\left(k\right)\right)
D_{q,1}^{h_1h_2}\left(\vect{P}_{\perp},\vect{R}_{\perp}\right)
\non\\
& = \int d^2 \kT d^2 \vect{P}_{\perp} \delta^2\left(\vect{P}_T - \vect{P}_{\perp}-z \kT\right)
 \non \\
& \times  \left(f_1^q\left(k\right)+\frac{[\bfS \times \bfk_T]_3}{M}f_{1T}^{ \perp q}\left(k\right)\right)
\\
& \times D_{q,1}^{h_1h_2}\left(\vect{P}_{\perp},\vect{R}_{T}-\frac{1}{2}(z_1-z_2) \kT\right)
\label{EQ_2H_SIG_SIV_R1},
}
where
\al{
D_{q,1}^{h_1h_2}\left(\vect{P}_{\perp},\vect{R}_{\perp}\right) \doteq
D_{q}^{h_1h_2}\left(\frac{1}{2}\vect{P}_{\perp}+\vect{R}_{\perp},\frac{1}{2}\vect{P}_{\perp}-\vect{R}_{\perp}\right).
\label{EQ_D1}
}
For the $\vect{P}_T$-integrated cross section now we have
\al{
\non
& \frac{d\sigma^{h_1h_2}}{d{\fS}\, d^2\vect{R}_{T}} = \int d^2 \kT d^2 \vect{P}_{\perp}
\left(f_1^q\left(k\right)+\frac{[\bfS \times \bfk_T]_3}{M}f_{1T}^{ \perp q}\left(k\right)\right)
\\
& \times D_{q,1}^{h_1h_2}\left(\vect{P}_{\perp},\vect{R}_{T}-\frac{1}{2}(z_1-z_2) \kT\right).
\label{EQ_2H_SIG_SIV_R1_INT}
}
We can see that with this choice of the relative transverse momentum the quark intrinsic $\kt$ appears also in the argument of DiFF and. Thus, in general, the Sivers effect in $\vect{R}_{T}$-dependent cross section will be nonzero for asymmetric ($z_1 \neq z_2$) pair production.

The model independent form of the unintegrated 2h SIDIS cross section that includes the Sivers effect in terms of structure functions is
\al{
\label{EQ_2H_X_SEC_RT}
&\frac{d\sigma^{h_1 h_2}}{d^2\vT\, d^2\vect{R}_T\,} = C(x,Q^2)x
\left[\sigma_U \vphantom{\frac{1}{1}} \right.
\\ \non
&
\left.+
S_T \left(\sigma_T\frac{P_T}{M}\sin(\vf_T-\fS) + \sigma_R\frac{R_T}{M}\sin(\vf_R-\fS)\right)\right],
}
where $\sigma_T = \frac{1}{2}\left(\sigma_1+\sigma_2\right)$,  $\sigma_R = \sigma_1-\sigma_2$. The explicit expressions for $\sigma_T$ and $\sigma_R$ with adopted parametrization of PDFs and DiFFs are given in~\cite{Kotzinian:2014gza}, where it is also shown that in general $\sigma_R \neq 0$ for asymmetric pair production.

 Having in mind the future experimental extractions of the Sivers asymmetries, it is useful to calculate the cross section in Eq.~(\ref{EQ_2H_X_SEC_RT}) after integrating over  the azimuthal angle of the relative or total transverse momentum, $\fR$ or $\fT$ respectively:
\al{
\label{EQ_2H_X_SEC_INT_R}
\frac{d\sigma^{h_1h_2}}{d^2\vT RdR} &= C(x, Q^2)
\left[
\sigma_{U,0}  \vphantom{\frac{1}{1}} \right.
\\ \non
&\left. +
S_T \left(\frac{P_T}{M}\sigma_{T,0}+\frac{R}{2M}\sigma_{R,1} \right)\sin(\vf_T-\fS)
\right],
\label{EQ_2H_X_SEC_INT_T}
\\
\frac{d\sigma^{h_1h_2} }{P_T dP_T d^2\vect{R}} &= C(x, Q^2)
\left[
 \sigma_{U,0}
\vphantom{\frac{1}{1}} \right.
\\ \non&
\left. +
S_T \left(\frac{P_T}{2M}\sigma_{T,1}+\frac{R}{M}\sigma_{R,0} \right)\sin(\vf_R-\fS)
\right],
}
where $\sigma_{U,i}, \sigma_{T,i}$ and $\sigma_{R,i}$ are the zeroth ($i=0$) and the first ($i=1$) harmonics of the $\cos(n(\vf_T-\vf_R))$ Fourier expansions of the corresponding structure functions. We see that both the $\sin(\fT-\fS)$ and $\sin(\fR-\fS)$ modulations have contributions from both the $\sigma_{T}$ and $\sigma_{R}$ unintegrated  cross section terms.

\section{Modified \LEPT (\MLEPT) results}
\label{SEC_MLEPTO}

\begin{figure}[tb]
\centering
\subfigure[] {
\includegraphics[width=\ImS]{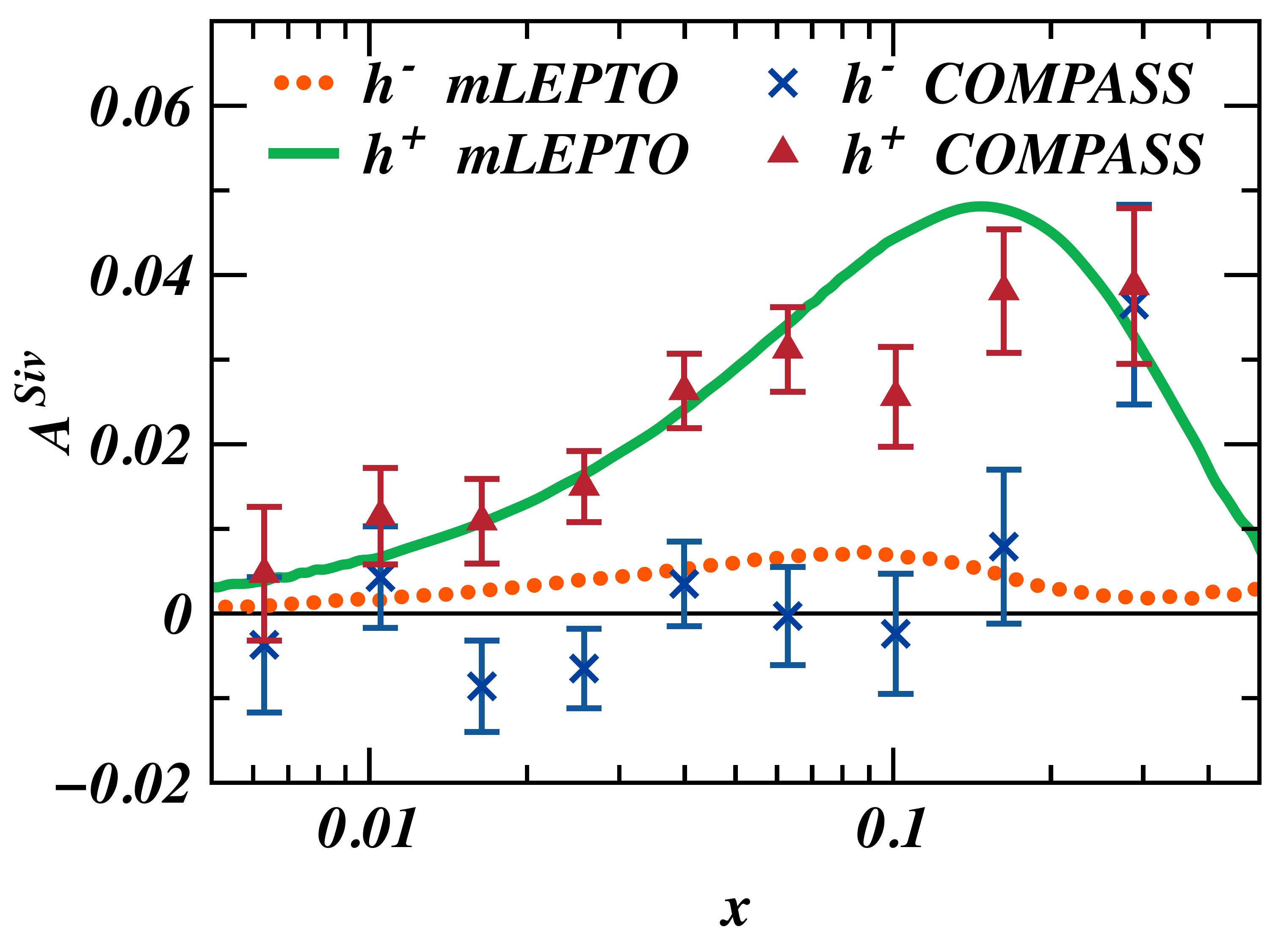}}
\\\vspace{-0.2cm}
\subfigure[] {
\includegraphics[width=\ImS]{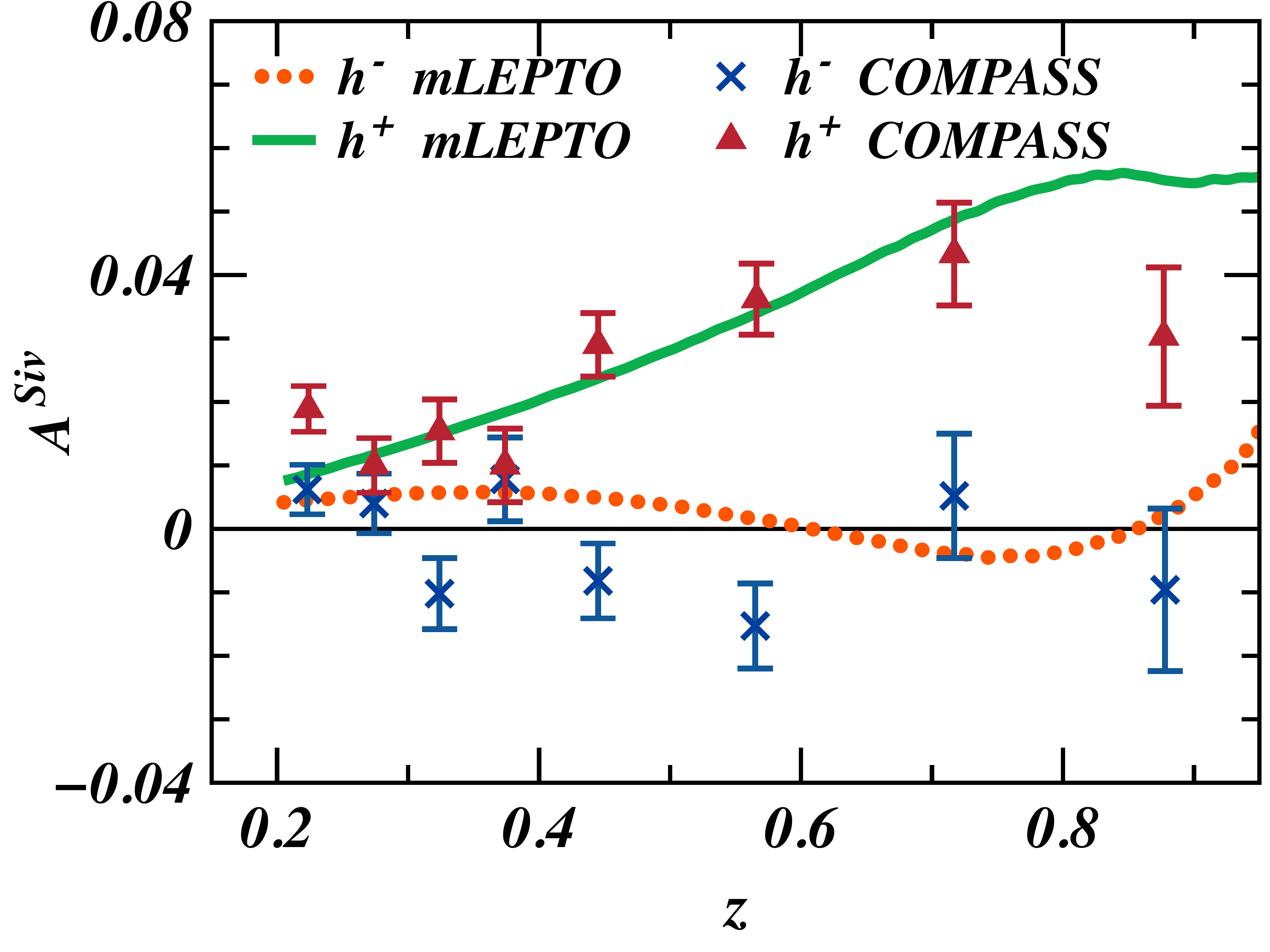}
}
\\\vspace{-0.2cm}
\subfigure[] {
\includegraphics[width=\ImS]{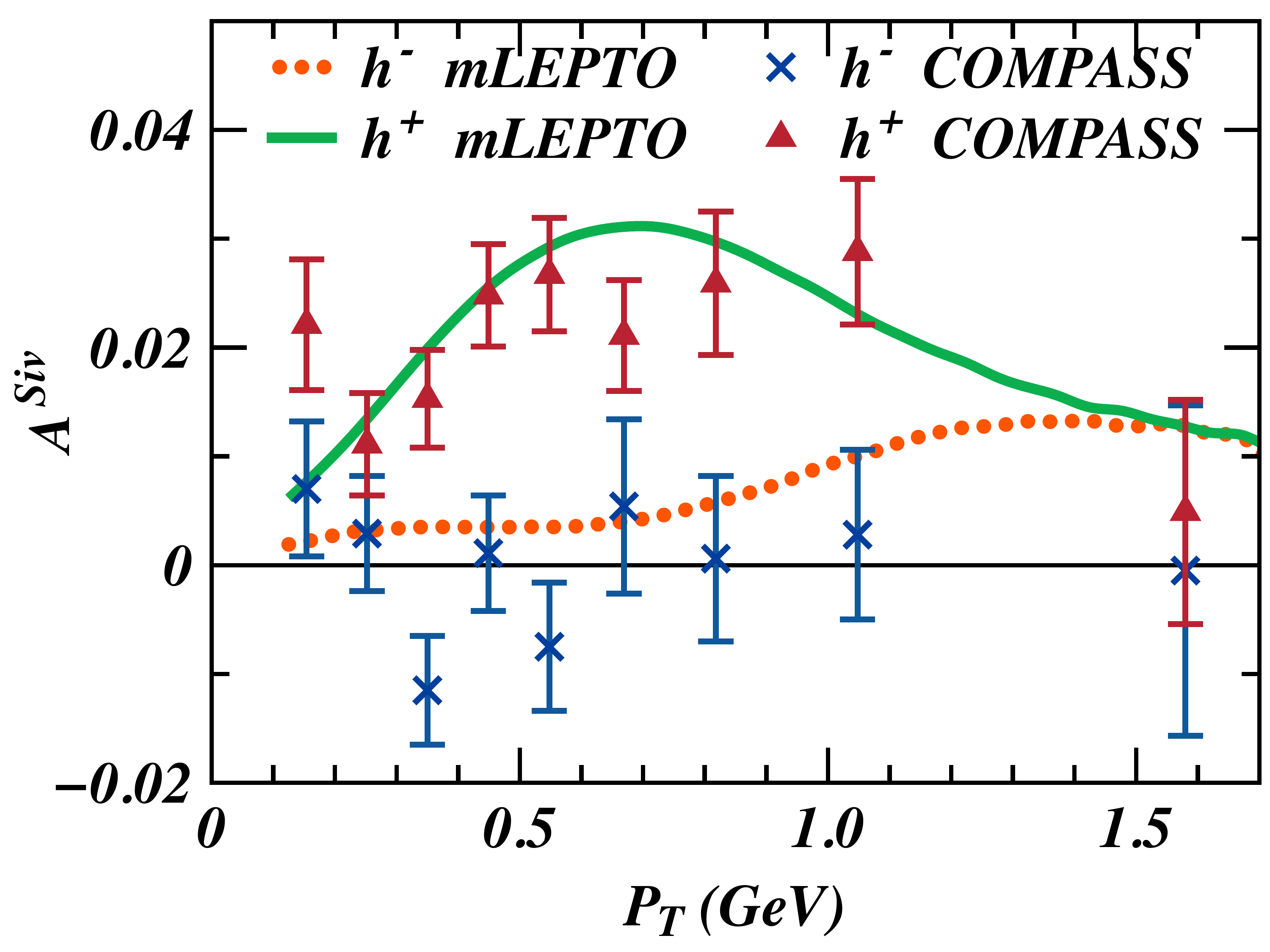}
}
\\\vspace{-0.2cm}
\caption{COMPASS results for Sivers asymmetry in a charged (triangles for positive and crosses for negative) hadron production off proton target, compared to those from \MLEPT (bands), for  $x$ (a), $z$ (b), and  $P_T$ (c) dependencies. The width of each band indicates the statistical accuracy of our simulations and does not include the uncertainties of the PDFs.}
\label{PLOT_SIV_1H}
\vspace{-1.3cm}
\end{figure}

 In this section we present numerical results obtained using a modified version (\MLEPT)~\cite{Kotzinian:2005zs,Kotzinian:2005zg} of  \LEPT~\cite{Ingelman:1996mq} unpolarized event generator. The Sivers effect is introduced as a modulation of the struck quark momentum's azimuthal angle according to the empirical parametrizations of Sivers function of Ref.~\cite{Anselmino:2005nn}, that were fitted to the experimental data~\cite{Airapetian:2009ae,Adolph:2012sp}.

In mLEPTO simulations we generated $10^{11}$ DIS events in the kinematical region of the the COMPASS experiment~\cite{Adolph:2012sp}, where $E_\mu=160~\Ge$. The following kinematic cuts were applied:  $Q^2>1~\Gs$, $0.1<y<0.9$, $0.03<x<0.7$, $W>5~\Ge$.  The single hadron SSAs calculated in this simulations then can be directly compared with those measured in Ref.~\cite{Adolph:2012sp}.

 We present the results of our \MLEPT simulations with only a single $\sin(\varphi-\varphi_S)$  modulation kept (see Eqs.~(\ref{EQ_2H_X_SEC_INT_R},\ref{EQ_2H_X_SEC_INT_T})), where $\varphi=\varphi_h$ for one hadron production, and $\varphi=\fR$ or $\varphi=\fT$ for two hadron production (see Eqs.~(\ref{EQ_2H_X_SEC_INT_R},\ref{EQ_2H_X_SEC_INT_T})). Then the cross section can be written as
\al{
d\sigma \propto \sigma_U+S_T \tilde{\sigma}_S \sin(\varphi-\varphi_S),
}
where $\sigma_U$ and $\tilde{\sigma}_S$ are the corresponding unpolarized and Sivers terms, and the asymmetry is defined as
\al{
A_{Siv}=\frac{\tilde{\sigma}_S}{\sigma_U}.
}
%

\begin{figure}[tb]
\centering
\includegraphics[width=\ImS]{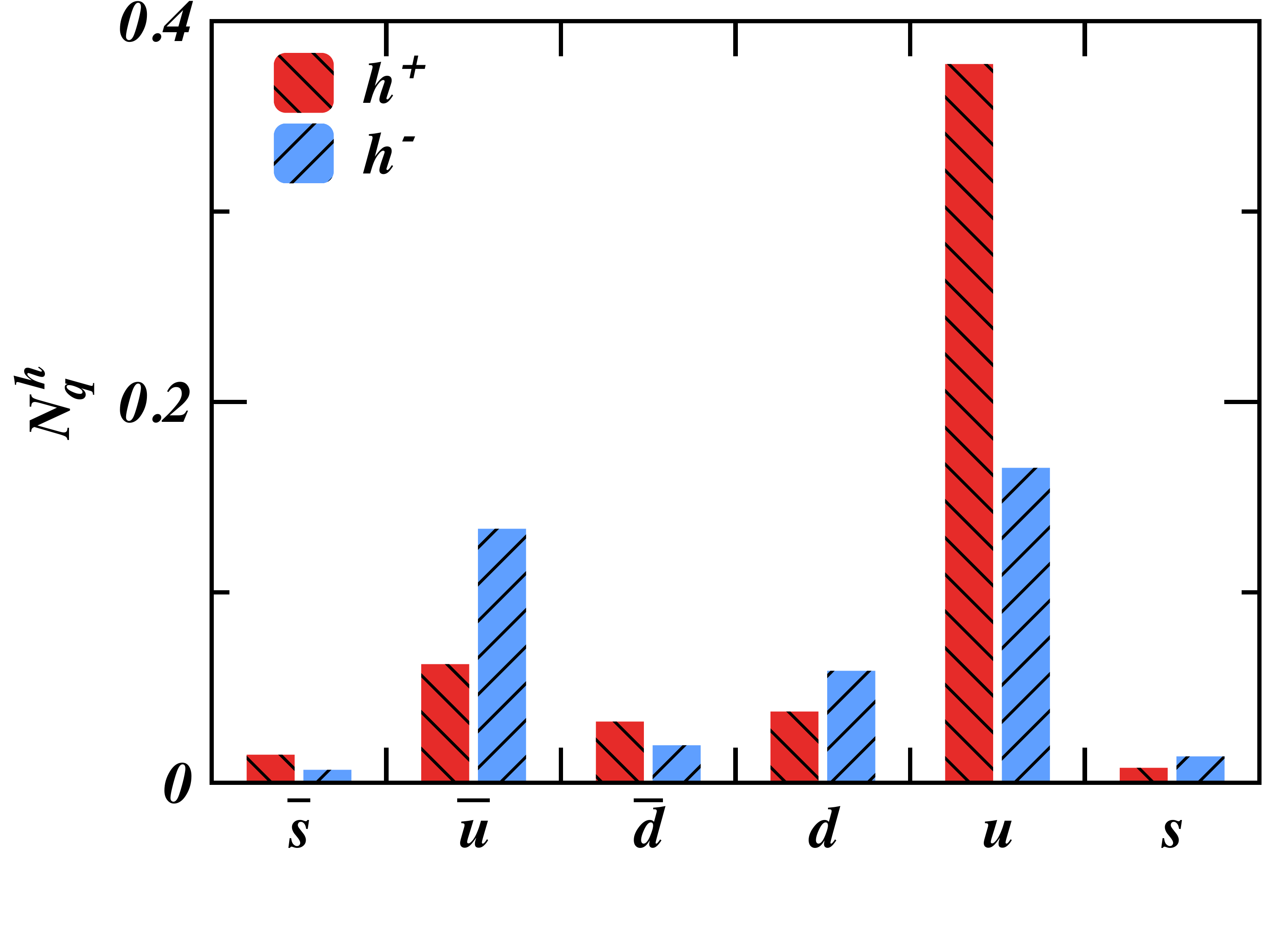}
%
\\\vspace{-0.2cm}
\caption{\MLEPT predictions for the relative rates for the flavor of the struck quark that produces positively (in red) and negatively (in blue) charged hadrons in MC events with all the relevant kinematical cuts.}
\label{PLOT_SIV_QUARK_ID}
\end{figure}
\begin{figure}[tb]
\centering
\subfigure[] {
\includegraphics[width=\ImS]{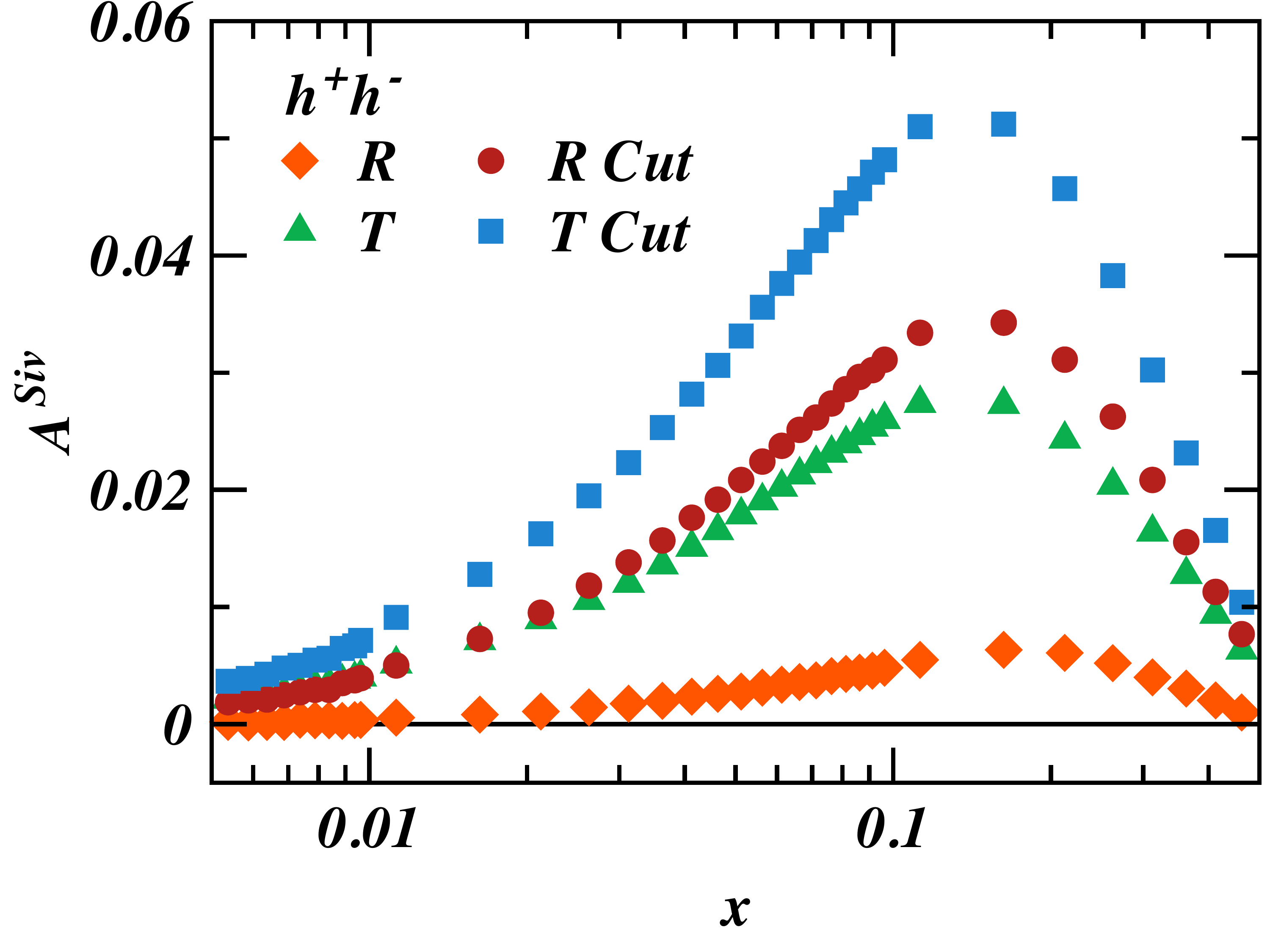}}
\\\vspace{-0.2cm}
\subfigure[] {
\includegraphics[width=\ImS]{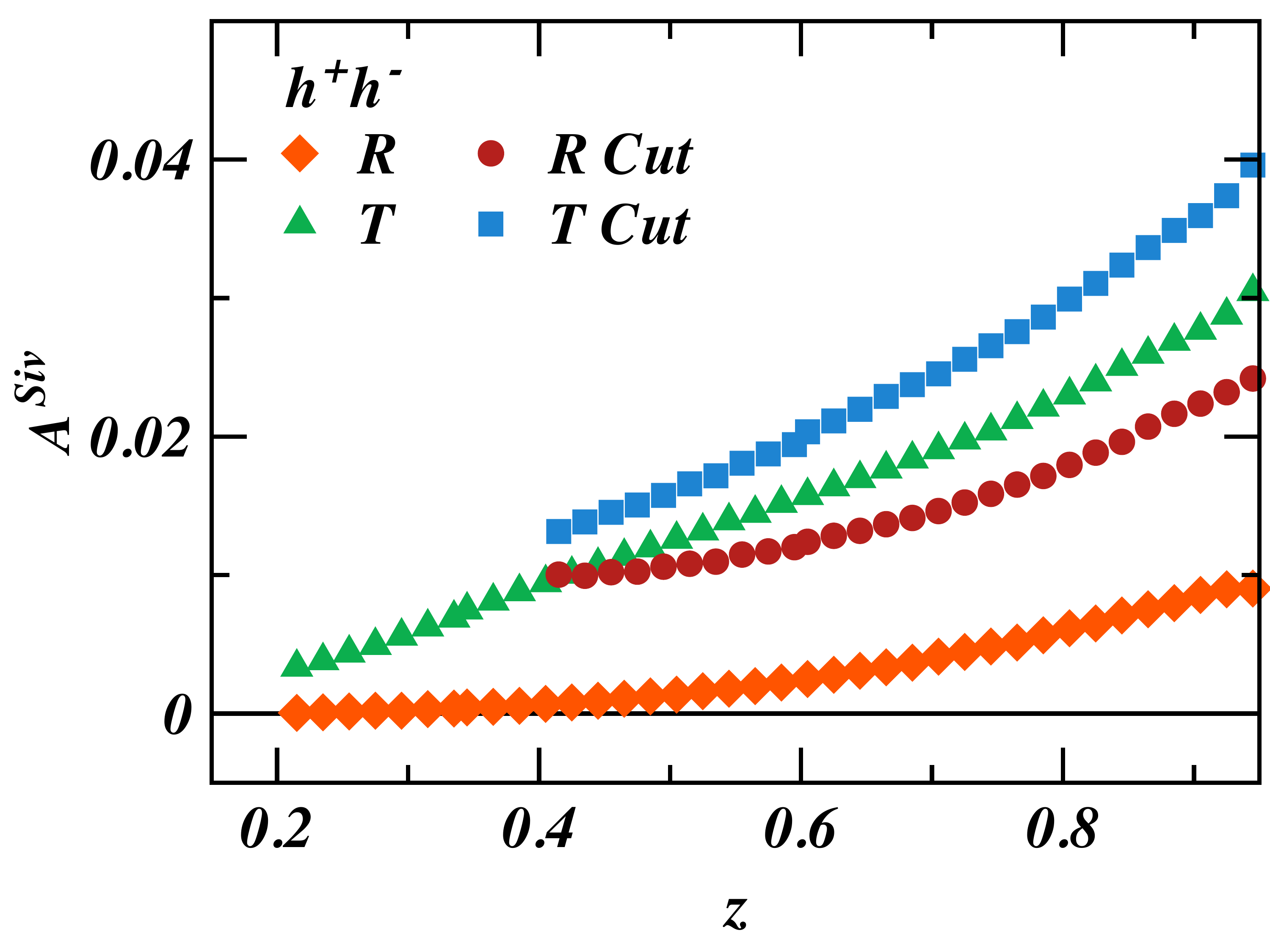}
}
\\\vspace{-0.2cm}
\caption{\MLEPT predictions for the dependence of the Sivers asymmetry on: (a) $x$, and (b) the total energy fraction $z$, in oppositely charged hadron pair production off proton target for both $\fR$ and $\fT$ asymmetries integrated over $\vect{P}_T$ and $\vect{R}$, respectively. The bands labeled "Cut" are the results with the additional cut on the positively charged hadron's momentum.}
\label{PLOT_SIV_2H_MIPL_X_Z}
\vspace{-1.2cm}
\end{figure}

 In Fig.~\ref{PLOT_SIV_1H} we compare \MLEPT results  for the SSAs in single hadron SIDIS production off transversely polarized proton target with the measurements by the COMPASS collaboration~\cite{Adolph:2012sp}, with the additional kinematic cuts $P_T>0.1~\Ge$ and $z>0.2$ used in the experiment. This allows us to validate \MLEPT by demonstrating that it provides a good description of the data used in extracting the parametrizations of Sivers PDFs.

 In \MLEPT we can determine the flavor of the struck quark in each SIDIS event. This proves to be a very useful information for understanding the underlying phenomenology of our results. In Fig.~\ref{PLOT_SIV_QUARK_ID}  the results for the relative rates for the struck quark flavor in events that produce a given type hadron satisfying the kinematic cuts imposed in our analysis are depicted. We see that the production of the positively charged hadrons, $h^+$,  is dominated by the $u$ quarks, while in production of the negatively charged hadrons, $h^-$,  both $u$ and $\bar{u}$ quark contribute almost equally, along with a significant input by the $d$ quark. We can easily interpret these by by examining the relative sizes of the quark PDFs and the fragmentation functions within the kinematical limits of the simulations.

 Our predictions for the dependence of the Sivers SSAs for a SIDIS production of a hadron pair with opposite charges $h^+h^-$ are presented in Fig.~\ref{PLOT_SIV_2H_MIPL_X_Z}, where in the pairs we choose $h^+$ as the first hadron and $h^-$ as the second hadron. We imposed the following kinematic cuts  on the momenta of the hadrons $P_{1(2)T}>0.1~\Ge$, $z_{1(2)}>0.1$. Here we extract the SSAs corresponding to both $\sin(\fT-\vf_S)$ and $\sin(\fR-\vf_S)$ modulations (labelled "T" and "R" on the plots). We note that  the SSAs, particularly those for the "R" modulations are significantly smaller than those for the single hadron case. This might make them difficult to measure experimentally, and thus might be harder to access experimentally. Fortunately, the SSAs can be enhanced by imposing asymmetric cuts, $z_1>0.3$ and $P_{T1}>0.3~\Ge$, (bands labeled "Cut" in Fig.~\ref{PLOT_SIV_2H_MIPL_X_Z}) on the momenta of the hadrons in a pair, as expected from the results in the previous Section.
\begin{figure}[tb]
\centering
\subfigure[] {
\includegraphics[width=\ImS]{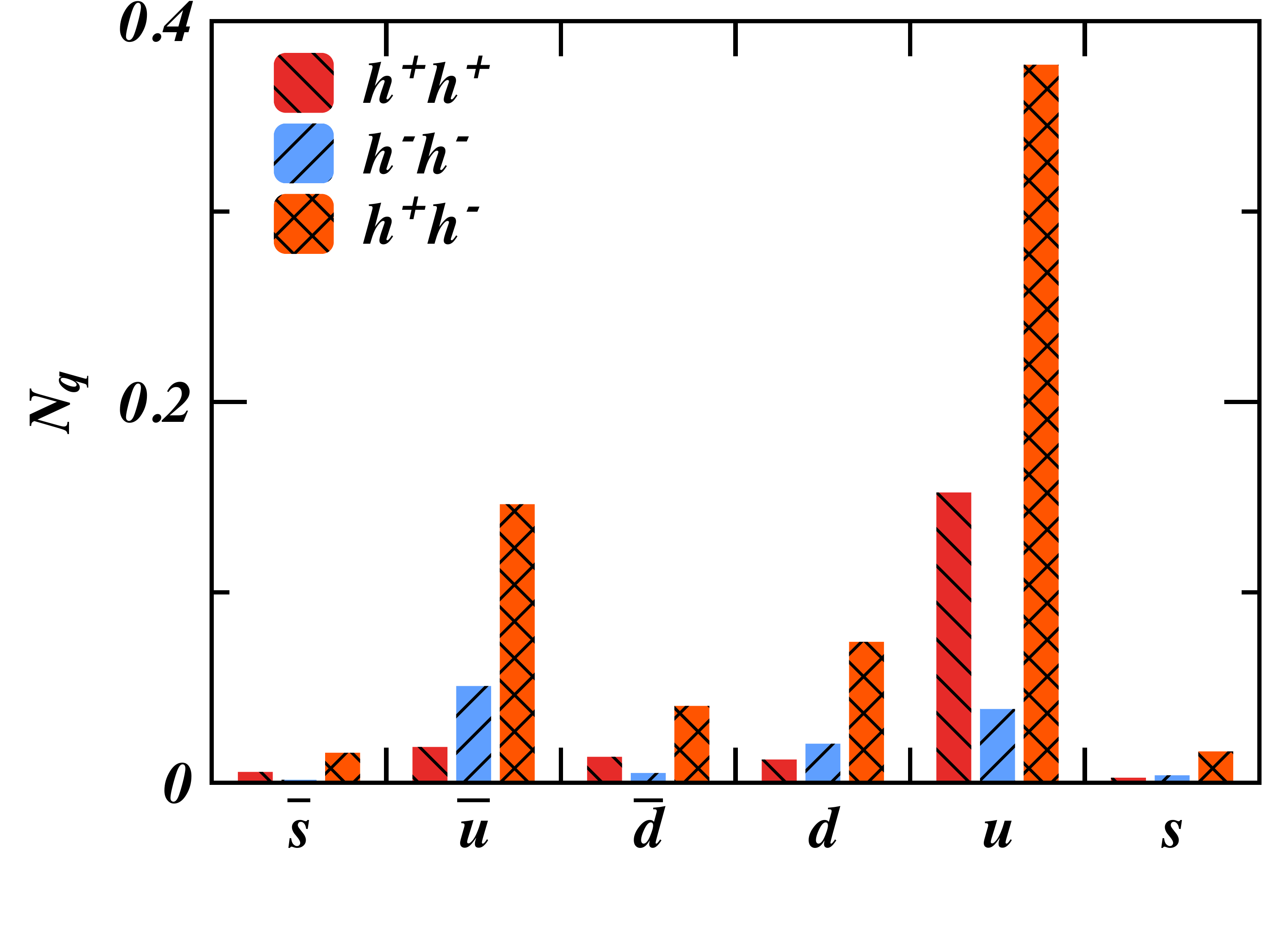}
}
\\\vspace{-0.2cm}
\subfigure[] {
\includegraphics[width=\ImS]{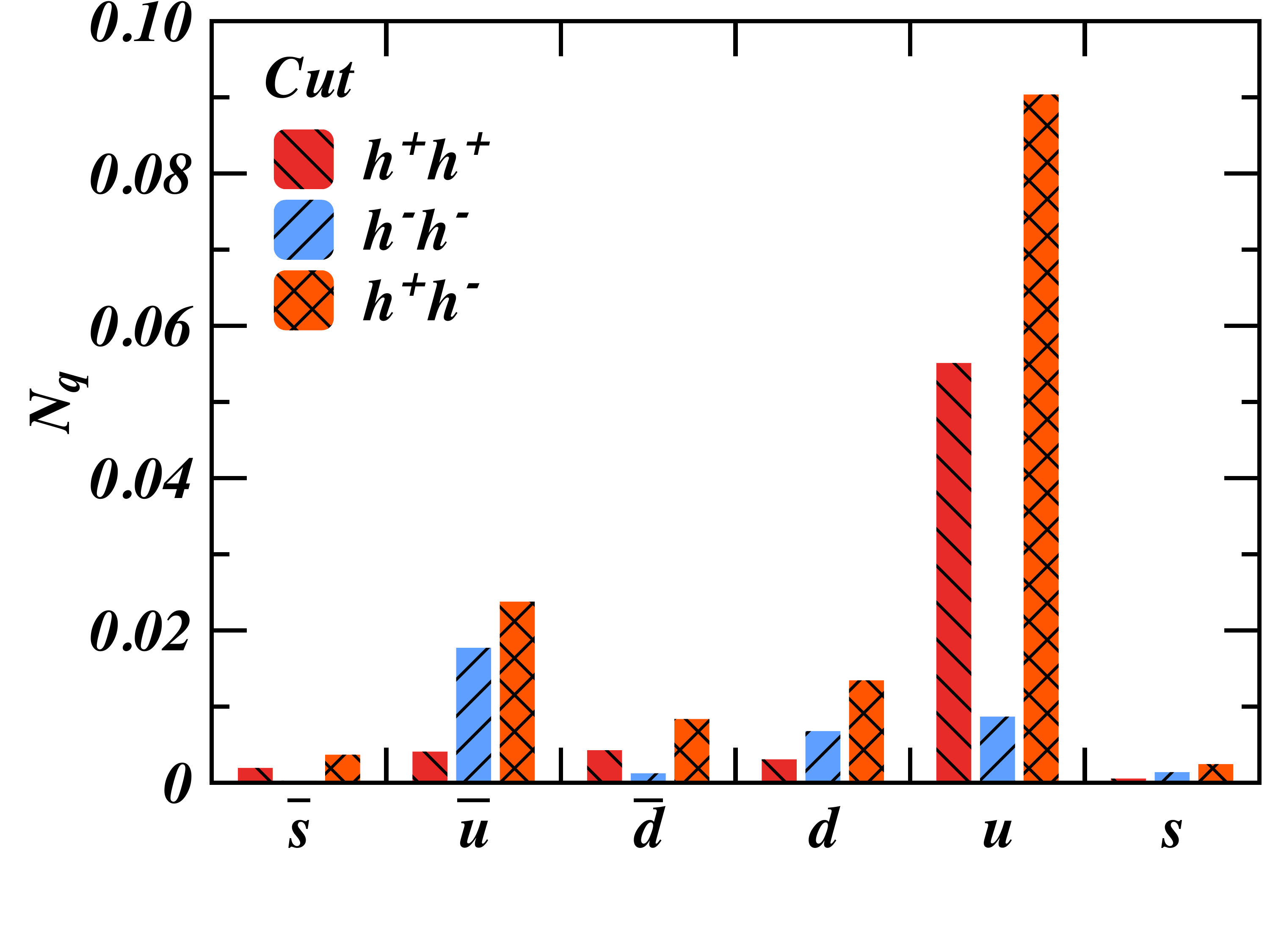}
}
\\\vspace{-0.2cm}
\caption{\MLEPT predictions for the relative rates for the flavor of the struck quark that produces a pair of both positively charged (in red), both negatively charged (in blue), and oppositely charged (in orange) hadrons in MC events with the relevant kinematical cuts (a), and additional asymmetric cuts on the momenta of the first hadron in the pair (b).}
\label{PLOT_SIV_QUARK_DIHAD_ID}
\vspace{-0.5cm}
\end{figure}

  We can also generate non zero "R" type SSAs for same-charged pairs by ordering the hadrons in the pairs according to their energy fraction: the hadron $h_1$ would be the one with the larger value of $z$ (i.e. $z$-order the hadrons in a pair). The SSAs for the pairs with the same charge involve convolutions of the Sivers function with a different set of dihadron fragmentation functions than those used in SSAs of oppositely charged hadrons.  These differences can be significant, analogous to the results of Ref.~\cite{Matevosyan:2013aka} for the NJL-jet calculations of dihadron fragmentation functions that depend on $z_h$ and $M_h$.  This expectation is confirmed by analyzing the relative rates of the different hadron pair production and their dependence on the struck quark's flavor from the plots in Fig.~\ref{PLOT_SIV_QUARK_DIHAD_ID}. Both for the  pairs with no additional cuts  depicted in Fig.~\ref{PLOT_SIV_QUARK_DIHAD_ID}(a) and for those with asymmetric cuts on the momenta of the hadrons in the pair in Fig.~\ref{PLOT_SIV_QUARK_DIHAD_ID}(b), the oppositely charged  pair production channel is predominant. Nevertheless,  $h^+h^+$ pairs, produced predominantly by the $u$ quark,  have a rate comparable to that of $h^+h^-$ pairs. On the other hand, the rate for $h^-h^-$ pairs is significantly smaller compared to the others, which impact strongly on the statistical errors for the extracted SSAs. This, compounded by the small values of the extracted asymmetries compared to other pairs, led us to omit the results for $h^-h^-$ pairs from this article.
\begin{figure}[tb]
\centering
\subfigure[] {
\includegraphics[width=\ImS]{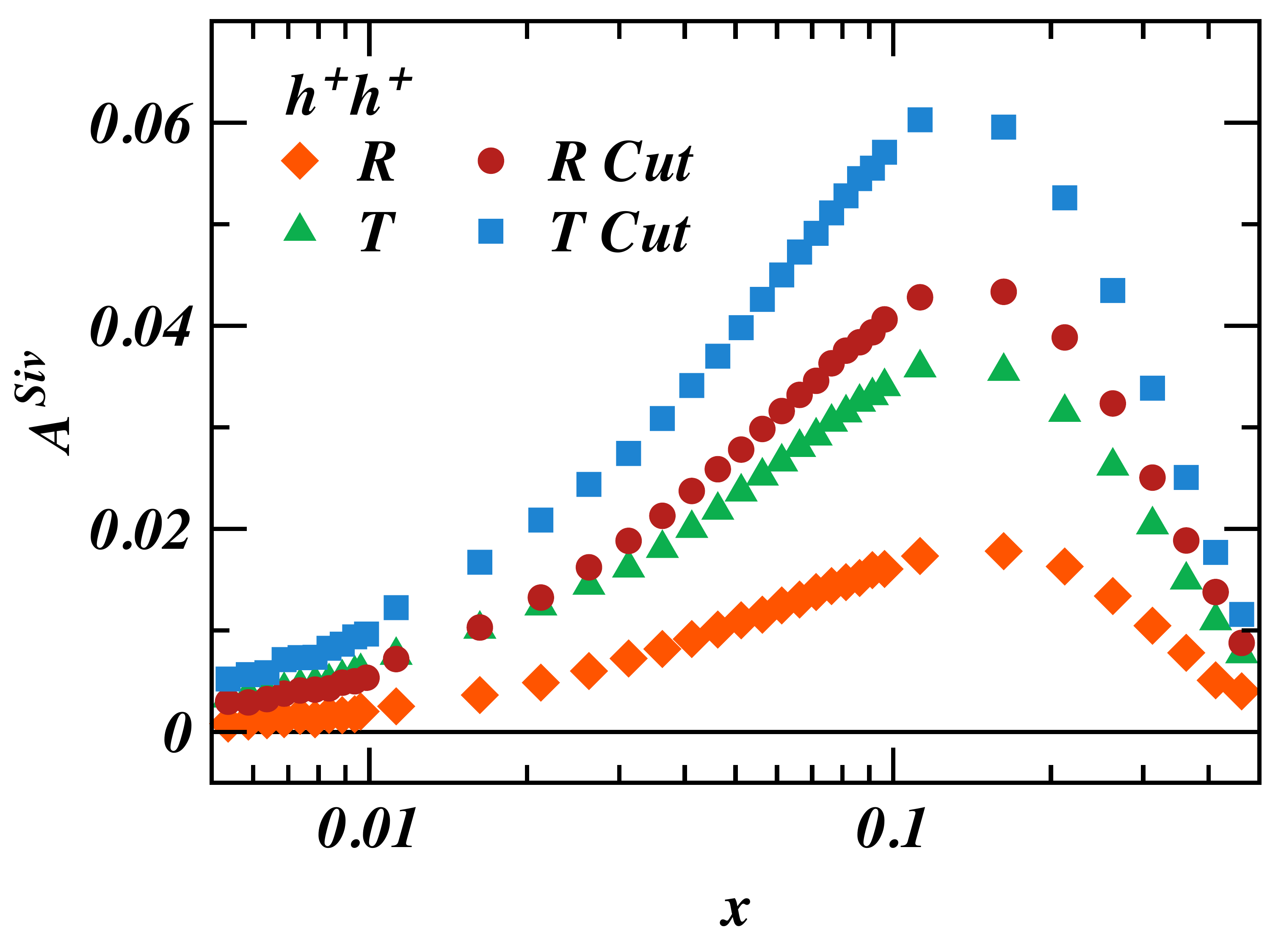}}
\\\vspace{-0.2cm}
\subfigure[] {
\includegraphics[width=\ImS]{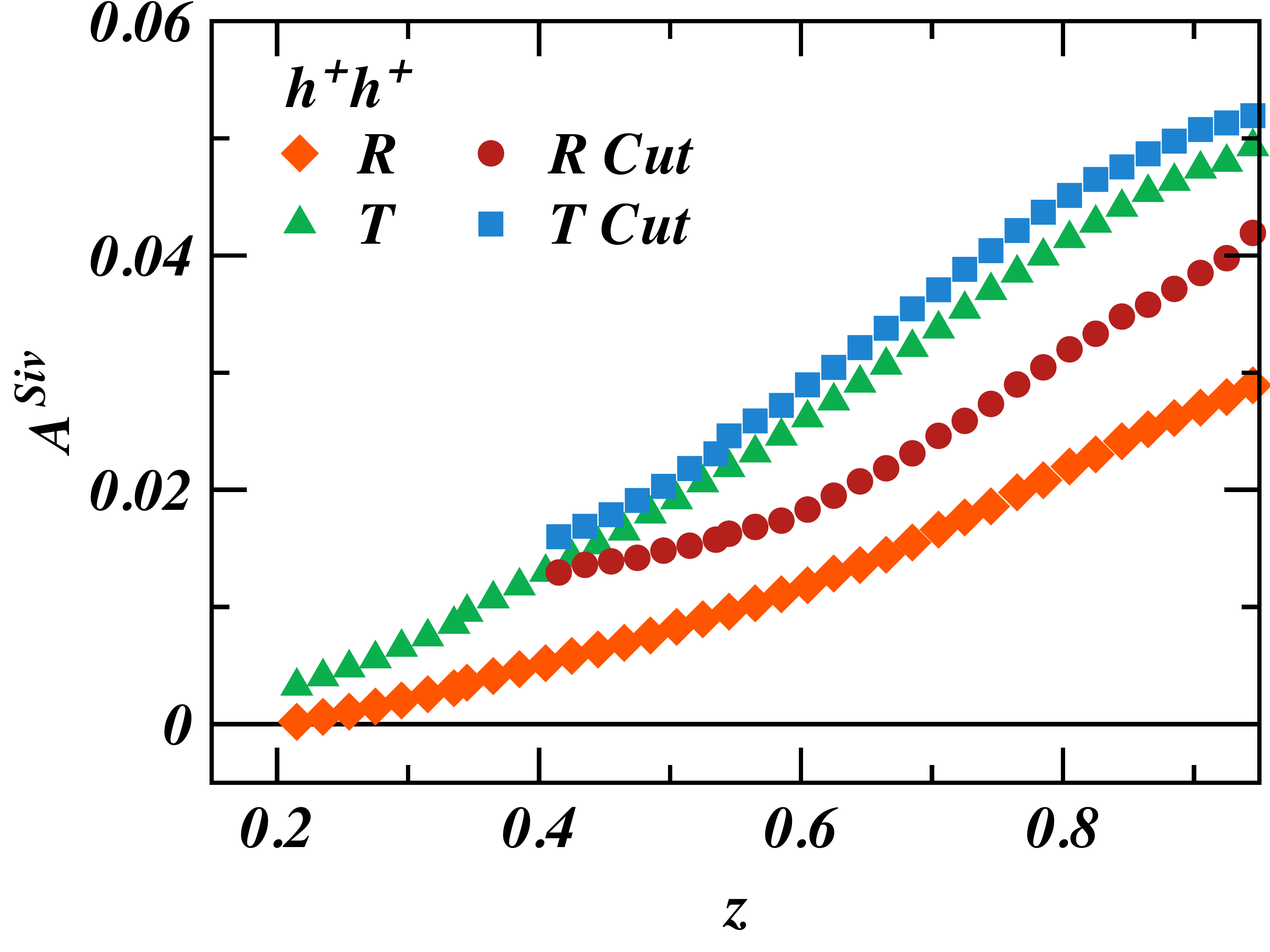}
}
\\\vspace{-0.2cm}
\caption{\MLEPT predictions for the dependence of the Sivers asymmetry on: (a) $x$, and (b) the total energy fraction $z$, in positively charged hadron pair production off a proton target for both $\fR$ and $\fT$ asymmetries integrated over $\vect{P}_T$ and $\vect{R}$, respectively. The bands labeled "Cut" are the results with the additional cut on the first hadron's momentum, as described in the text.}
\label{PLOT_SIV_2H_PLPL_X_Z}
\vspace{-0.8cm}
\end{figure}

The results for $h^+h^+$ pairs are depicted in the Figs.~\ref{PLOT_SIV_2H_PLPL_X_Z}, where we see that the SSAs for the $h^+h^+$ pairs are larger than those for $h^+h^-$ pairs. This will counteract the negative impact of the larger statistical errors stemming for the smaller rate of the $h^+h^+$ pairs compared to $h^+h^-$ in the experimental  measurements.

\section{Conclusions}
\label{SEC_CONC}

  In our recent work~\cite{Kotzinian:2014lsa,Kotzinian:2014gza}, we proposed a new approach for measuring the Sivers function in the 2h SIDIS process. We presented the results for the relevant cross section calculation that involve the Sivers PDF, with two non-vanishing SSAs with respect to $\sin(\fR-\fS)$ and $\sin(\fT-\fS)$, respectively. In contrast, the leading twist expression for the cross-section of the two-hadron SIDIS presented in Ref.~\cite{Bianconi:1999cd} with different choice of the relative transverse momentum contains only a $\sin(\fT-\fS)$ modulation term. The $\sin(\phi_{R,P}-\fS)$ modulation term is absent there, as well as in the subleading twist expression of Ref.~\cite{Bacchetta:2003vn} when integrated over the total transverse momentum of the pair.

  Here we have presented the detailed explanation of origin of the 'unexpected' nonzero $\sin(\fR-\fS)$ modulation in 2h SIDIS. We showed that, using the leading order approximations the  $\sin(\fR-\fS)$  type modulations with the choice of the relative transverse momenta of Refs.~\cite{Bianconi:1999cd,Bacchetta:2003vn} should indeed vanish, while they should be in general nonzero with our definition of $\fR$.  The actual measurement of sine modulations in terms of above two different choices of the relative transverse momentum can serve as a good test of the applicability of factorization  formalism and various leading order approximation in 2h SIDIS at the energy scale of that experiment.

We conclude that the magnitude of Sivers SSAs in the dihadron SIDIS process should be comparable to those for the single hadron SIDIS. Thus the experimental measurements of the Sivers SSAs for various hadron pairs  using the data collected at COMPASS and the future SIDIS experiments, such as those planned at JLAB12 and EIC, are feasible and would contribute a large amount of new information for extracting the Sivers PDFs.

\bibliography{fragment}

\begin{thebibliography}{18}

\bibitem{Kotzinian:2014lsa}
A.~Kotzinian, H.H. Matevosyan, A.W. Thomas, to be published in Phys.Rev.Lett.
  (2014), \texttt{1403.5562}

\bibitem{Kotzinian:2014gza}
A.~Kotzinian, H.H. Matevosyan, A.W. Thomas (2014), \texttt{1405.5059}

\bibitem{Sivers:1989cc}
D.W. Sivers, Phys.Rev. \textbf{D41}, 83 (1990)

\bibitem{Airapetian:2009ae}
A.~Airapetian et~al. (HERMES Collaboration), Phys.Rev.Lett. \textbf{103},
  152002 (2009), \texttt{0906.3918}

\bibitem{Adolph:2012sp}
C.~Adolph et~al. (COMPASS Collaboration), Phys.Lett. \textbf{B717}, 383 (2012),
  \texttt{1205.5122}

\bibitem{Qian:2011py}
X.~Qian et~al. (Jefferson Lab Hall A Collaboration), Phys.Rev.Lett.
  \textbf{107}, 072003 (2011), \texttt{1106.0363}

\bibitem{Anselmino:2005nn}
M.~Anselmino, M.~Boglione, U.~D'Alesio, A.~Kotzinian, F.~Murgia et~al.,
  Phys.Rev. \textbf{D71}, 074006 (2005), \texttt{hep-ph/0501196}

\bibitem{Airapetian:2008sk}
A.~Airapetian et~al. (HERMES Collaboration), JHEP \textbf{0806}, 017 (2008),
  \texttt{0803.2367}

\bibitem{Vossen:2011fk}
A.~Vossen et~al. (Belle Collaboration), Phys.Rev.Lett. \textbf{107}, 072004
  (2011), \texttt{1104.2425}

\bibitem{Adolph:2014fjw}
C.~Adolph et~al. (COMPASS Collaboration), Phys.Lett. \textbf{B736}, 124 (2014),
  \texttt{1401.7873}

\bibitem{Artru:1995zu}
X.~Artru, J.C. Collins, Z.Phys. \textbf{C69}, 277 (1996),
  \texttt{hep-ph/9504220}

\bibitem{Bianconi:1999cd}
A.~Bianconi, S.~Boffi, R.~Jakob, M.~Radici, Phys.Rev. \textbf{D62}, 034008
  (2000), \texttt{hep-ph/9907475}

\bibitem{Bacchetta:2003vn}
A.~Bacchetta, M.~Radici, Phys.Rev. \textbf{D69}, 074026 (2004),
  \texttt{hep-ph/0311173}

\bibitem{Arneodo:1986yc}
M.~Arneodo et~al. (European Muon Collaboration), Z.Phys. \textbf{C31}, 333
  (1986)

\bibitem{Kotzinian:2005zs}
A.~Kotzinian, Proceedings of Transversity 2005, pp. 228--235 (2005),
  \texttt{hep-ph/0510359}

\bibitem{Kotzinian:2005zg}
A.~Kotzinian (2005), \texttt{hep-ph/0504081}

\bibitem{Ingelman:1996mq}
G.~Ingelman, A.~Edin, J.~Rathsman, Comput.Phys.Commun. \textbf{101}, 108
  (1997), \texttt{hep-ph/9605286}

\bibitem{Matevosyan:2013aka}
H.H. Matevosyan, A.W. Thomas, W.~Bentz, Phys.Rev. \textbf{D88}, 094022 (2013),
  \texttt{1310.1917}

\end{thebibliography}

\end{document}